\documentclass[usegraphicx,usenatbib]{mn2e}

\usepackage{graphicx}
\usepackage[usenames,dvips]{color}

\usepackage{longtable}

\def\arcsec{$\,^{\prime\prime}$~}
\def\arcmin{$\,^\prime$~}
\def\deg{$^{\circ}$~}

\newcommand{\bfalpha}{\mbox{\boldmath $\alpha$}} 
\newcommand{\bfdelta}{\mbox{\boldmath $\delta$}}

\newcommand{\lsim }{{\lower0.8ex\hbox{$\buildrel <\over\sim$}}}
\newcommand{\gsim }{{\lower0.8ex\hbox{$\buildrel >\over\sim$}}}

\newcommand{\chandra}{\emph{Chandra}}
\newcommand{\B}{\mbox{$B_{435}$}}
\newcommand{\R}{\mbox{$R_{625}$}}
\newcommand{\ha}{\mbox{H$\alpha$}}
\newcommand{\br}{\mbox{$\B\!-\!\R$}}
\newcommand{\hr}{\mbox{$\ha\!-\!\R$}}

\def\Chandra{${\it Chandra}$}

\def\AE{${\it AE}$}

\def\simge{\mathrel{%
   \rlap{\raise 0.511ex \hbox{$>$}}{\lower 0.511ex \hbox{$\sim$}}}}
\def\simle{\mathrel{
   \rlap{\raise 0.511ex \hbox{$<$}}{\lower 0.511ex \hbox{$\sim$}}}}

\newcommand{\Msun}{\ifmmode {M_{\odot}}\else${M_{\odot}}$\fi}
\newcommand{\Lsun}{\ifmmode {L_{\odot}}\else${L_{\odot}}$\fi}
\newcommand{\Rsun}{\ifmmode {R_{\odot}}\else${R_{\odot}}$\fi}

\begin{document}

\title [Faint Millisecond Pulsars of NGC 6752]{A {\it Chandra} Look at the X-ray Faint Millisecond Pulsars in the Globular Cluster NGC 6752}
\author[Forestell et al.]{L.~M.~Forestell,$^{1}$ C.~O.~Heinke,$^{1}$\thanks{Ingenuity New Faculty; heinke@ualberta.ca} H.~N. Cohn,$^{2}$ P.~M.~Lugger,$^{2}$ G.~R.~Sivakoff,$^{1}$ \newauthor S.~Bogdanov,$^{3}$ A.~M.~Cool,$^{4}$, J.~Anderson$^{5}$ \smallskip\\
$^{1}${Physics Dept., University of Alberta, 4-183 CCIS, Edmonton, AB T6G 2G7, Canada}\\
$^{2}${Department of Astronomy, Indiana University, 727 E. Third St., Bloomington, IN 47405, USA }\\
$^{3}${Columbia Astrophysics Laboratory, Columbia University, 550 West 120th Street, New York, NY 10027, USA}\\
$^{4}${Department of Physics and Astronomy, San Francisco State University, 1600 Holloway Avenue, San Francisco, CA 94132, USA}\\
$^{5}${Space Telescope Science Institute, Baltimore, MD 21218, USA} }
\maketitle 

\begin{abstract}
We combine new and archival \Chandra\ observations of the globular cluster NGC 6752 to create a deeper X-ray source list, and study the faint radio millisecond pulsars (MSPs) of this cluster.  We detect four of the five MSPs in NGC 6752, and present evidence for emission from the fifth.  The X-rays from these MSPs are consistent with thermal emission from the neutron star surfaces, with significantly higher fitted blackbody temperatures than other globular cluster MSPs (though we cannot rule out contamination by nonthermal emission or other X-ray sources).  NGC 6752 E is one of the lowest-$L_X$ MSPs known, with $L_X$(0.3-8 keV)=$1.0^{+0.9}_{-0.5}\times10^{30}$ ergs s$^{-1}$. We check for optical counterparts of the three isolated MSPs in the core using new HST ACS images, finding no plausible counterparts, which is consistent with their lack of binary companions.  We compile measurements of $L_X$ and spindown power for radio MSPs from the literature, including errors where feasible.  We find no evidence that isolated MSPs have lower $L_X$ than MSPs in binary systems, omitting binary MSPs showing emission from intrabinary wind shocks.  We find weak evidence for an inverse correlation between the estimated temperature of the MSP X-rays and the known MSP spin period, consistent with the predicted shrinking of the MSP polar cap size with increasing spin period.

\end{abstract}



\section{Introduction}\label{s:intro}
The cores of globular clusters (GCs) may reach high stellar densities, up to $10^6$ times that of local space, that can lead to significant dynamical interactions, producing compact binary systems that can engage in mass transfer. 
 Thus, GCs are very efficient at producing interacting binary stars, including low-mass X-ray binaries \citep[LMXBs,][]{Clark75}, radio millisecond pulsars \citep[MSPs,][]{Johnston92}, and cataclysmic variables \citep[CVs,][]{Pooley03}. 
MSPs are the progeny of LMXB evolution, in which a low mass star transfers angular momentum to a neutron star (NS), spinning up the rotational period of the NS to millisecond timescales \citep{Bhattacharya91,Papitto13}.  

MSPs can produce both thermal and nonthermal X-rays \citep{Becker02, Zavlin02, Zavlin07}.  The nonthermal radiation (dominant in the MSPs with the highest spindown power, \.{E}) is attributed to the pulsar magnetosphere, is generally highly beamed (and thus sharply pulsed), and typically described by a power-law with a photon index $\sim$1.1-1.2  \citep{Becker99,Zavlin07}. 
 The thermal radiation is blackbody-like radiation from a portion of the NS surface around the magnetic poles, heated by a flow of relativistic particles in the pulsar magnetosphere to $\sim$1 MK \citep{Harding02}.  The X-ray spectra and rotation-induced pulsations of the nearby MSPs that exhibit thermal radiation are well-described by hydrogen atmosphere models \citep{Zavlin98,Bogdanov07,Bogdanov09}.  X-ray observations of a large sample of MSPs allow study of how the thermal radiation from MSPs relates to other pulsar parameters \citep{Kargaltsev12}.  Due to the high density of MSPs in GCs, and the well-known distances and reddening to GCs, GCs are ideal targets for such studies.  

NGC 6752 is a GC located at a distance of 
$4.0 \pm 0.2$ kpc \citep[][2010 revision]{Harris96}.\footnote{http://physwww.physics.mcmaster.ca/$\sim$harris/mwgc.dat} 
Its reddening of $E_{B-V}=0.046$ \citep{Gratton05} can be converted to a neutral gas column of $N_H=3.2\times10^{20}$ cm$^{-2}$ using the relation of \citet{Guver09}. 
The center of the cluster has been measured, using {\it Hubble Space Telescope (HST)}  images, to be at (J2000) $19^h10^m52^s.11$,~-59\deg 59\arcmin 04.4\arcsec (\cite{Goldsbury2010}). We adopt a core radius of 10.2\arcsec, and half-mass radius of 1.91\arcmin \citep[][2010 revision]{Harris96}, though the central parts of the surface brightness profile are poorly described by a single King model \citep[see, e.g.,][]{Thomson12}.

The cluster was first detected at X-ray wavelengths by \citet{Grindlay93} using the {\it ROSAT} satellite. Deeper {\it ROSAT} studies identified multiple X-ray sources within the cluster \citep{Johnston94,Verbunt00b}, and two CVs were identified in HST images at the positions of two X-ray sources \citep{Bailyn96}.  
\citet{Pooley02a} used the \Chandra\ X-ray Observatory to resolve the cluster emission into 19 X-ray sources within the half-mass radius, and used {\it HST} and {\it Australian Telescope Compact Array} radio images to confirm the two counterpart suggestions by Bailyn et al. and identify 6-9 more CVs, 1-2 chromospherically active binaries, and 1-3 background galaxies.  

 Five MSPs have been discovered in the cluster \citep{D'Amico02}, three of which lie within the core radius and show extreme line-of-sight accelerations indicative of a high mass density in the cluster core.  One pulsar (MSP A) lies 3.3 half-mass radii from the cluster center, suggesting that the pulsar either has been ejected (perhaps by an encounter with a massive black hole, or binary black hole, \citealt{Colpi02}), or is not associated with the cluster \citep{Bassa06}.  Four of the five MSPs are isolated, with only MSP A being in a binary system with an optically identified helium white dwarf companion \citep{Ferraro03,Bassa03}.  
\citet{D'Amico02} note that MSP D matches Pooley et al's CX11, which was identified by Pooley et al. as a CV or galaxy, based on their suggested optical counterpart (see below).  \citet{D'Amico02} also identify X-ray emission from MSP C, which lies outside the half-mass radius, and tentatively suggest X-ray emission from MSP B.  

We have obtained a new \Chandra\ observation, and combined it with the archival 2000 \Chandra\ observation to produce a deeper image of NGC 6752 and create a larger source catalog.  In this paper, we describe our X-ray analysis and the new source catalog, and focus on the X-ray properties of the MSPs in NGC 6752. In particular, we clearly identify X-ray emission from four MSPs, and find less certain evidence for X-ray emission from the fifth (MSP E).
 A companion paper, Lugger et al.\ (in prep) identifies optical counterparts for our extended X-ray source catalog using newly acquired {\it HST Advanced Camera for Surveys (ACS)} data.

\section{Observations and Analysis}
The globular cluster NGC 6752 was observed twice with the \Chandra~ACIS-S detector at the aimpoint. The first observation, described by \citet{Pooley02a}, was taken on 2000 May 15  (ObsID 948), lasting 29.85 ks. 
The second observation was taken on 2006 Feb.\ 10 (ObsID 6612), for a total time of 38.45 ks. Both observations placed the core of the GC on the S3 CCD, which has increased sensitivity to low energy X-rays, of the ACIS-S detector. Observation 948 was performed in timed-exposure, Faint mode, which uses a 3$\times$3 pixel island for grade classification of each event. Observation 6612 used the timed-exposure, Very Faint mode, utilizing a 5$\times$5 pixel island for superior grade classification and rejection of cosmic rays. For this observation, we selected an offset and roll angle to ensure that MSP A fell on the S3 chip, as its position was not covered by any chip in the first observation.

\begin{figure*}
\begin{minipage}{126mm}
\hspace{-2cm}
  \includegraphics[width=6.5in]{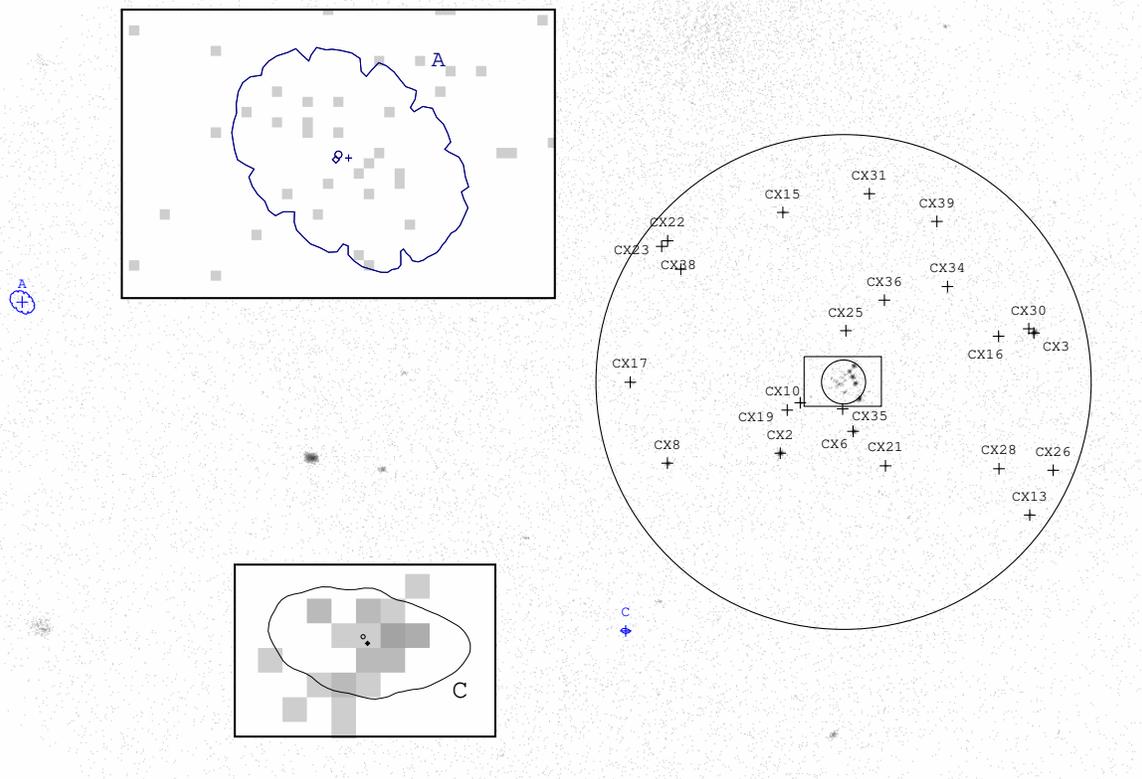}
  \caption{Combined \Chandra~data of NGC 6752. Circles indicate the half-mass and core radii of 1.91\arcmin and 10.2\arcsec, respectively. The rectangle indicates the smaller core region, that is detailed in Figure~\ref{inner_catalog}. The locations of the outer MSPs (A and C) are identified, with an ellipse indicating the extraction region for A. {\it Insets:} Zoom of the 0.3-2 keV images in the neighbourhood of MSPs A and C.  The extraction regions enclosing 90\% of the expected \Chandra\ PSF at the radio position of MSP A, and the detected position of MSP C, are indicated, as are the radio positions (crosses), and the data centroid positions (diamond) and PSF-correlation positions (circle) from \AE. }
  \label{outer_catalog}
\end{minipage}
\end{figure*}

\subsection{Data Reduction}
We reduced the data using CIAO version 4.4\footnote{http://cxc.cfa.harvard.edu/ciao/}, following standard CIAO science threads\footnote{http://cxc.harvard.edu/ciao/threads/index.html}. We limited the energy range to 0.3-10 keV, within which the ACIS CCDs are calibrated. We only extracted events from  the S3 CCD, which covered the cluster out to its half-mass radius in both observations. 
We further cleaned the data using the {\it deflare}\footnote{http://cxc.harvard.edu/ciao4.4/threads/flare/} CIAO process to remove any background flares in the ACIS datasets, so that the level 2 event file would be suitable for spectral extractions. 
The final good time intervals for the observations were 27.78 ks and 38.20 ks for the 948 and 6612 observations, respectively.
We combined the data (for imaging purposes) after matching the astrometry of the later observation to the earlier one. 
We created exposure maps and aspect histograms for the S3 CCD, and mask files and aspect files covering the time range of the observations, for use with the ACIS-EXTRACT (AE) algorithms discussed in the next section.

\subsection{Source Detection}
We detected sources using two detection algorithms, CIAO's {\it wavdetect} algorithm \citep{Freeman02}\footnote{http://cxc.harvard.edu/ciao/threads/wavdetect/}, and the independent {\it pwdetect} algorithm \citep{Damiani97} \footnote{http://www.astropa.unipa.it/progetti\_ricerca/PWDetect/}. 
We have found that {\it wavdetect} is efficient and highly reliable in detecting sources across wide fields, while {\it pwdetect} is more capable of detecting faint sources close to bright sources (e.g., in the cores of globular clusters).  For {\it wavdetect}, we created images in the 0.3-2 keV and 0.3-7 keV energy bands, using scales of 1.0, 1.4, and 2.0 pixels, with a source detection significance threshold of $10^{-6}$, which should result in one false detection per ACIS chip.  (We chose not to use larger source detection scales, since we were primarily interested in point sources near the aimpoint, where permitting larger detection scales can merge multiple faint sources together.) For {\it pwdetect}, we used the same images, using wavelet scales from 0.5'' to 2.0'', and a final detection threshold of 5.1$\sigma$, which should also result in one false detection per ACIS chip.  Except for the likely X-ray counterparts to MSPs A and C (see below), we only report sources within the cluster half-mass radius (18\% of the area of the S3 chip), and 
we expect less than one false detection even given four detection runs.

After creating a combined source list, the catalog was further refined using the \AE\ package\footnote{http://www2.astro.psu.edu/xray/acis/acis\_analysis.html}, detailed by \cite{Broos10,Broos12}. Initial extractions of spectra and background for each source were performed, merging data from the observations in 2000 and 2006. \AE~was then used to refine the position of each object in the catalog, calculating the centroid of the data within a preliminary extraction region,
as recommended by \citet{Broos10}.  If the probability of the extracted counts being produced by fluctuations in the background was above a threshold value of 10\% 
(as calculated in \citealt{Weisskopf07}), the source was removed from the catalog, the positions refined, and the process repeated until the catalog no longer required pruning.

\begin{figure}
  \centering
  \includegraphics[width=1\columnwidth]{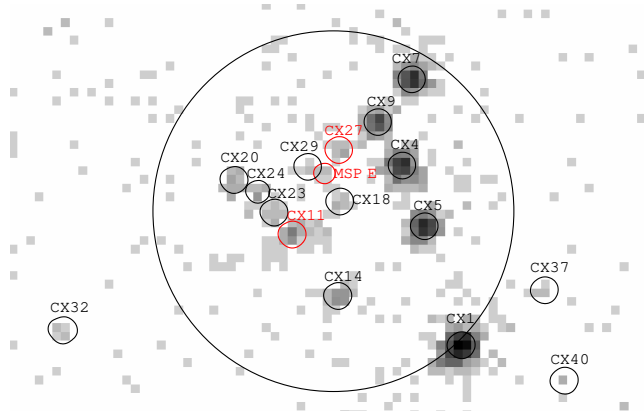}
  \caption{Combined \Chandra~data of NGC 6752. Shown is the inner region of Figure~\ref{outer_catalog}, with catalog sources identified, and the cluster core radius indicated. The extraction regions (defined by ACIS\_EXTRACT) for our catalog X-ray sources are plotted. Red extraction regions indicate the MSPs; CX11 corresponds to MSP D, while CX27 corresponds to MSP B, and MSP E is formally undetected, though X-ray emission is visible at its location.}
  \label{inner_catalog}
\end{figure}

This pruning left us with 39 X-ray sources detected within the cluster half-mass radius. 
 Figure~\ref{outer_catalog} shows the sources found in the outer region of the cluster, while Figure~\ref{inner_catalog} details the sources within the core. Circles indicating the half-mass and core radius are shown for clarity \citep{Harris96}.  The source positions are ordered by average flux (from highest to lowest) and labelled accordingly in 
Table \ref{Catalog_Parameters}.  We retain the numbering scheme of \citet{Pooley02a} for their detected sources (except their CX12, which we resolve into three sources), and number additional sources by decreasing 0.5-8 keV flux.

We performed a careful study of the positions of the five radio MSPs detected by \cite{D'Amico02}, three of which (B, C, and D) were detected by our detection algorithms. 
Figure~\ref{outer_catalog} shows the locations of the outer MSPs A and C. 
MSP A lies 6.39\arcmin from the center of the cluster \cite{D'Amico02},  or 3.3 half-mass radii from the center, and its PSF is therefore quite broad.  Since MSP A was not detected by our standard detection algorithms (see details below), we added a source at its position. \AE's catalog pruning process did not remove this source, indicating that it is detected by our observations. MSP C lies 1.4 half-mass radii from the cluster center, and is clearly detected (as previously reported by \citealt{D'Amico02}). 
Figure~\ref{inner_catalog} shows the core sources determined in our catalog, and identifies the location of the MSPs. MSPs B and D are in excellent agreement ($<0.3$'') with our detected sources CX27 and CX11, respectively.
We checked the astrometry of our X-ray observations by using the secure X-ray detections of MSPs C, B, and D to match the X-ray astrometry to the radio positions and astrometric frame, giving a net shift of the X-ray positions of -0.035'' in RA and +0.155'' in Dec. 

On the other hand, MSP E is neither clearly detected nor clearly undetected. Lying $\sim$1\arcsec from our detected source CX29, MSP E is in a region near the cluster center that contains emission from multiple sources (Figure~\ref{inner_catalog}).  
Inspection of the region suggests that there is a faint source located at the position of MSP E that was not detected due to the close proximity of CX29, CX27 (MSP B), and CX18.  We extract data from the location of this source as for the other MSPs, but the lack of a clear detection means that we cannot be certain that the X-ray emission within our extraction region is from MSP E.

\subsection{Extraction and Photometry}
Following source position refinement, the \AE\ package was used to extract final source and background spectra for the catalog X-ray sources, which include MSPs B, C, and D, and for the extraction regions at the positions of MSPs A and E. 
The sources were extracted multiple times, with each extraction optimized for a different reason. One extraction was done to check the source position, one to check whether each source could be explained as a background fluctuation (to weed out spurious sources), and the final extraction was optimized for photometric and spectroscopic analysis. 

For each source, events were selected from within a region that encompasses 90\% of the PSF  centered on each catalog position, or a region of reduced size if the sources were too crowded. Background extractions were constructed using the \AE\ $\it{better\_backgrounds}$ algorithm, and effective area files (ARFs) and response matrices (RMFs) were constructed for each source. Background extractions  included at least 100 counts, and sample pixels from areas outside all source extraction regions, selecting the background region to accurately assess the local background due to neighboring point sources as well as the instrumental background. 

Background subtracted photometry was calculated in several bands. The number of counts for each catalog source and each MSP was determined in the soft (0.5--1.5 keV), hard (1.5--6 keV) and broad (0.5--8 keV) bands, for comparison with previous work \citep[e.g.][]{Bogdanov06}. The total flux in the broad band was also calculated, using XSPEC version 12.7\footnote{http://heasarc.gsfc.nasa.gov/docs/xanadu/xspec/}. 
For all sources with less than 100 total counts, hereafter the combined faint sample, we applied the XSPEC MEKAL model, accounting for Galactic absorption with the TBABS model, to the combined spectrum. We choose the MEKAL model since we expect these faint sources to be dominated by chromospherically active binaries and cataclysmic variables, both of which have X-ray spectra well-represented by MEKAL models \citep[e.g.][]{Heinke05a}.
We computed a countrate-to-flux conversion from this, and used it to calculate fluxes for the fainter sources.  
For the brightest nine 
 catalog sources, each spectrum was fit independently with several models.  Bremsstrahlung models were found to be perfectly adequate, as expected for thermal plasma at high temperatures).  
Since these sources have mostly been identified, through their optical counterparts, as cataclysmic variables \citep[CVs,][]{Pooley02a}, we expect hard X-ray spectra consistent with  bremsstrahlung emission. To model the spectra of the MSPs, a blackbody model was used, as found appropriate for most X-ray faint MSPs \citep{Bogdanov06,Bogdanov11}. The calculated fluxes were converted to unabsorbed luminosities in the 0.5-8keV range (Table~\ref{Catalog_Parameters}).

\begin{figure}
  \includegraphics[width=1\columnwidth]{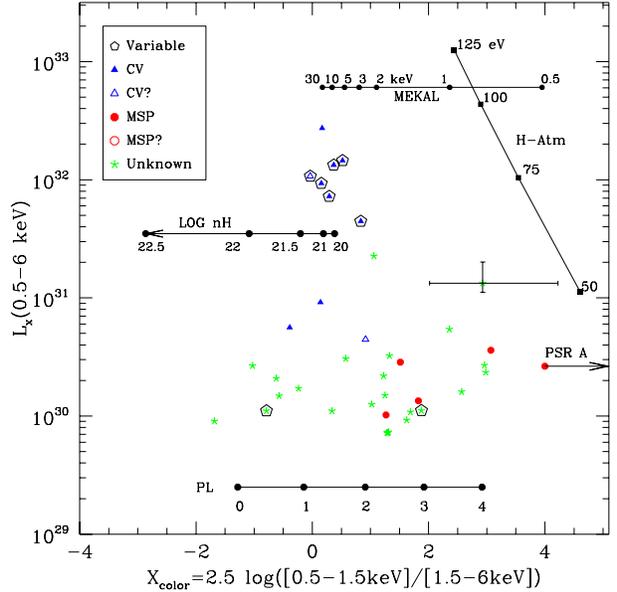}
  \caption{X-ray CMD for the cluster NGC 6752. Plotted is X-ray luminosity (broad band) against hardness (increasing to the left). CVs, MSPs, and unknown X-ray sources are plotted independently. 
Variable sources (identified by whether a Kolmogorov-Smirnov test finds the distribution of photon arrival times inconsistent with a constant) are boxed. 
Plotted for comparison are the neutron star hydrogen atmosphere model (assuming a 10 km NS and cluster absorption), power law model, MEKAL model, and a 10 keV MEKAL model with varying $N_H$.  (Apart from the neutron star model, the models have arbitrary normalization.) }
  \label{CMD}
\end{figure}

We created an X-ray color magnitude diagram (CMD) of the cluster, plotting X-ray hardness ([1.5-6 keV counts]/[0.5-1.5 keV counts]) versus the inferred 0.5-6 keV luminosity (Figure~\ref{CMD}). Also plotted are theoretical lines for  MEKAL, power law, and NS hydrogen atmosphere models in XSPEC with varying temperatures or photon indices.  A fixed 10 keV MEKAL model with varying N$_H$ values is plotted to indicate the effects on color of increasing $N_H$.  Comparison of this X-ray CMD with those of other clusters with numerous optical counterparts (e.g., 47 Tuc, \citealt{Grindlay01a,Heinke05a}; NGC 6397, \citealt{Grindlay01b}; M4, \citealt{Bassa04}; $\omega$ Cen, \citealt{Haggard10}; also see  \citealt{Pooley06}) shows the same principal features.  CVs are concentrated at an X-ray color near 0 (hard spectra consistent with power-laws of photon index $\Gamma=$1-2, or thermal plasma with  $kT>$2 keV), with $L_X$ between a few $10^{30}$ and a few $10^{32}$ ergs/s.  The radio MSPs are softer (colors consistent with power-law photon indices $\Gamma>$2) with $L_X<4\times10^{30}$ ergs s$^{-1}$, making them less luminous on average than those in 47 Tuc. Comparison of the positions of chromospherically active binaries will require additional optical counterpart identifications in NGC 6752 (\citealt{Pooley02a,Thomson12}; see Lugger et al., in prep.).

\subsection{Spectral Fitting of MSPs}

\begin{figure}
  \centering
  \includegraphics[width=1\columnwidth]{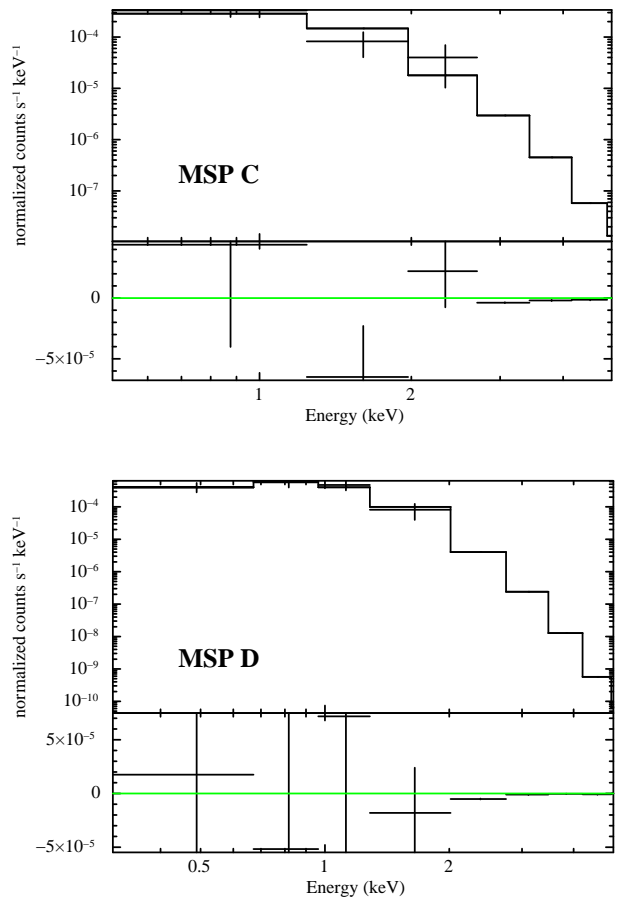}
  \caption{Spectral results for the two brightest MSPs, C and D. The spectra were modelled with an absorbed blackbody (top panels are data and model, lower panels are residuals). The models were fitted with unbinned data, but presented here binned to at least 30 channels/bin to improve readability. The residuals for both models are also given.
}
  \label{BB_MODELS}
\end{figure}

For the X-ray faint MSPs, we used the C-statistic, to perform spectral fitting with few photons \citep{Cash79}. In place of the reduced $\chi^2$ statistic to test whether a model is a good fit, we use the ``goodness 1000'' command in XSPEC, which generates 1000 Monte Carlo simulations of the chosen model to see what fraction have a lower fitting statistic than the actual data (rejecting models with, say, goodness $>$95\%).  The poor statistics from the MSPs also forced us to freeze the hydrogen column density to that of the cluster  ($N_H=2.2\times10^{20}$ cm$^{-2}$, using the TBABS absorption model, \citealt{Wilms00}), as it could not be reasonably constrained by spectral fits.   Freezing the $N_H$ to the cluster value is reasonable, as none of the MSPs possess companions that are losing mass (4 are single, the other has a white dwarf companion).  Thus, we do not expect extra gas to be associated with these systems. 
We fit the MSPs first to the XSPEC blackbody model BBODYRAD, hereafter referred to as BB,  providing constraints on the effective radii of the X-ray emitting regions (Table~\ref{MSP_Parameters}).  Example spectral fits are illustrated in Figure~\ref{BB_MODELS}.  We compare the radii, temperature, and luminosities of MSPs in NGC 6752 to those in the clusters 47 Tuc and NGC 6397 \citep{Bogdanov06,Bogdanov10}.  Figure~\ref{Lx_T} compares luminosity versus temperature, while Figure~\ref{T_R} compares temperature versus radii.  We include source CX29 (modelled with a BB spectrum), the nearest detected source to MSP E, for comparison, as some of the photons from CX29 may have leaked into the extraction radius of MSP E.

We also fitted the spectra with a NS hydrogen atmosphere model (NSATMOS, \citealt{Heinke06a}).
  The NS mass and radius were fixed to 1.4 \Msun\ and 10 km respectively, and the distance to 4.0 kpc, while the normalization was left free (physically interpreted as a portion of the surface radiating). 
The results of both model fits for the temperature, radius, 
and luminosity are given in Table \ref{BB_NSATMOS}. 

\section{Results}

\subsection{MSP A}
\begin{figure}
  \centering
  \includegraphics[width=\columnwidth]{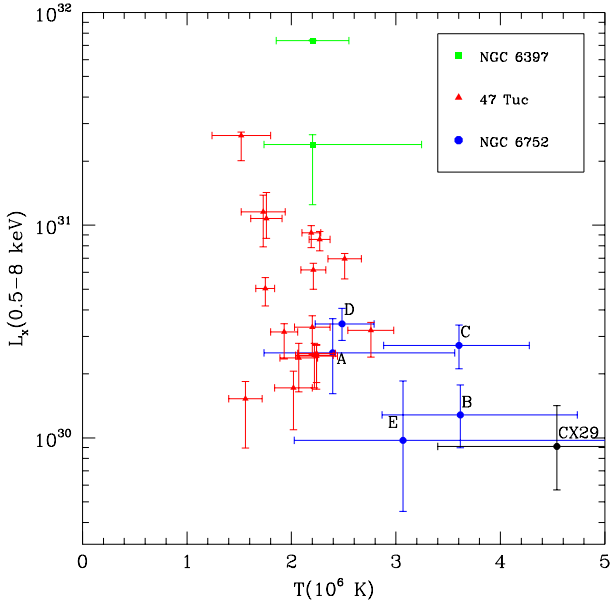}
  \caption{Broad band luminosity vs. temperature for NGC 6752 MSPs, and for those in 47 Tuc and NGC 6397. The temperatures and radii are taken from BB models of the MSPs.}
  \label{Lx_T}
\end{figure}

\begin{figure}
  \centering
  \includegraphics[width=\columnwidth]{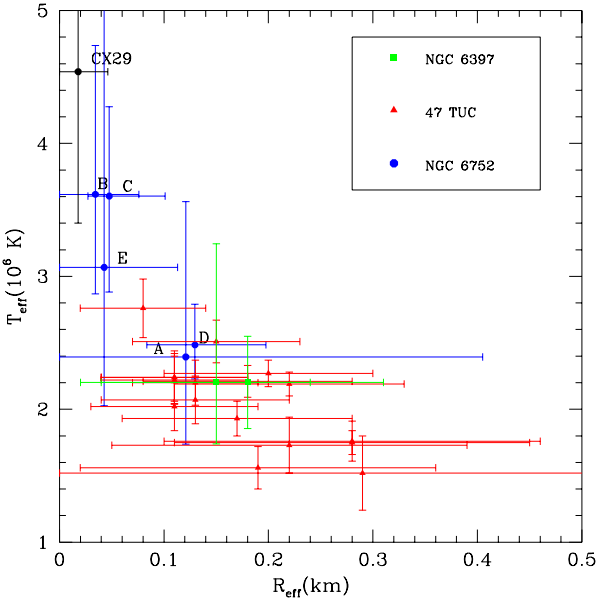}
  \caption{Temperature vs. radius for NGC 6752 MSPs, and for those in 47 Tuc and NGC 6397. The temperatures are taken from BB models of the MSPs.}
  \label{T_R}
\end{figure}

Because of the positioning of the \Chandra~ACIS-S array in the 2000 observation, MSP A was only observed in the 2006 observation. This MSP lies 3.3 half-mass radii away from the center of the cluster, and was not detected by our standard source detection algorithms (see Figure \ref{outer_catalog}).  However, we were able to obtain a detection using a reduced significance threshold ($10^{-5}$ vs. $10^{-6}$ in {\it wavdetect}) and appropriately large scales for the extended PSF at this position, giving a {\it wavdetect} source significance of 3.3.
Only 22 (12) counts, including an estimated 8 (2) background counts, were extracted from this region in the 0.5-8 (0.5-1.5) keV energy range.  The probability of these counts being due only to background fluctuations was low, $5.3\times10^{-5}$ ($8\times10^{-6}$), as computed by \AE\ for the 0.5-8 (0.5-1.5) keV band.
The region including 90\% of the PSF is unusually large (15\arcsec~$\times$10\arcsec), due to the large off-axis angle. This likely explains why neither {\it wavdetect} nor {\it pwdetect} originally detected this source.  
 The \AE\ source position estimates also agree reasonably well with the radio MSP position,  with the farthest less than 1\arcsec away from the extraction point. 
The blackbody fit yields a temperature of $T_{BB}=0.21^{+0.10}_{-0.06}$keV (Table~\ref{MSP_Parameters}). An  unabsorbed flux of $1.3\times10^{-15}$ ergs $s^{-1}$ cm$^{-2}$ was determined, corresponding to a luminosity of $(2.5^{+1.1}_{-0.9})\times10^{30}$ ergs $s^{-1}$.  The blackbody temperature and X-ray luminosity are consistent with the range of those of MSPs in other clusters (Figures \ref{Lx_T} and \ref{T_R}).


\subsection{MSP B}
MSP B's position coincides closely with the X-ray source CX27 (Figure \ref{inner_catalog}).
It is one of the fainter detected sources, with only 11 counts (one likely background). 
The blackbody fit gave a temperature of $T_{BB}=0.31^{+0.10}_{-0.07}$keV, and the luminosity was determined to be $(1.3^{+0.5}_{-0.4})\times10^{30}$ ergs s$^{-1}$.  This is an unusually low X-ray luminosity for an MSP, and an unusually high temperature for the thermal spectrum of an MSP.  Note that if the spectrum is contaminated by nonthermal radiation or other X-ray sources, the true thermal X-ray luminosity will be even lower.

\subsection{MSP C}
This source is located well outside of the crowded core, making the extraction, and the positional and source validity estimates, fairly robust (Figure \ref{outer_catalog}) . The source is 2.70\arcmin from the center of the cluster, or 1.4 half-mass radii. 
The isolated location of the MSP also provides confidence in our spectral modelling, as there are no nearby sources to cause confusion in the spectra. 
Modelling the MSP with a blackbody, we found $T_{BB}=0.31^{+0.06}_{-0.06}$keV, providing a luminosity of $(2.7^{+0.7}_{-0.6})\times10^{30}$ ergs s$^{-1}$. This is perhaps the clearest example of an unusually high blackbody temperature when compared to MSPs studied in other clusters (see Figures \ref{Lx_T} and \ref{T_R}). 

\subsection{MSP D}
This source is our brightest MSP, although it still contains only 33 counts. 
The fitted temperature is in better agreement with the temperatures from MSPs in other clusters, at $T_{BB}=0.21^{+0.03}_{-0.02}$ keV. The corresponding $L_X$ is $(3.4^{+0.6}_{-0.6})\times10^{30}$ ergs $s^{-1}$, similar to the average of MSPs seen in clusters like 47 Tuc. 

\subsection{MSP E}
The final MSP in the cluster is the least luminous, with only 6 counts in our extraction region. 
The emission appears reasonably centered on the position of MSP E, though it is not detected by our detection algorithms, likely due to crowding (Figure \ref{inner_catalog}).
Source CX29 from our catalog is the closest detected X-ray source, at 1.1'' from the position of MSP E.  If CX29's X-ray source position were incorrect, it is possible that the emission at the position of MSP E could be attributed to a source between the positions of MSP E and the putative CX29 position (we show evidence against this possibility in the next section). 
We attempted to model the extracted counts from the pulsar catalog position, using the C-statistic in XSPEC. For the blackbody model, this yielded $T_{BB}=0.27^{+0.19}_{-0.09}$keV, while the luminosity was computed to be $(1.0^{+0.9}_{-0.5})\times10^{30}$ erg s$^{-1}$, the lowest luminosity of any MSP known in the cluster.

\section{Optical Counterpart Search}

Due to the X-ray crowding in the core of this GC, a critical question is whether the X-ray emission from the positions of the MSPs in the core (B, D, and E) is due to those MSPs, or to other sources.  We address this question by searching for optical counterparts to the X-ray sources nearest our radio MSP positions in new, deep {\it HST} data.  Since these MSPs are isolated NSs, we expect essentially no optical emission from them, and thus an optical counterpart showing blue colors or \ha\ excess would indicate the presence of a cataclysmic variable or chromospherically active binary star, which could produce some of the X-ray emission.

A complete analysis of the optical counterparts to the X-ray sources
in NGC~6752 is reported in \citet[][in prep]{Lugger13}.  Here we briefly describe
the key steps in our analysis.  The analysis is based on deep
\emph{HST} ACS/WFC imaging of NGC~6752 in F435W (\B), F625W (\R), and
F658N (\ha) from the GO-12254 dataset (PI: Cool).  Multiple dithered
frames were combined using the STSDAS routine \emph{astrodrizzle} and
plate solutions relative to the ICRS were computed for the resulting
mosaic images using approximately 600 astrometric standards from the
USNO UCAC3 catalog.  The photometry of individual images was performed
using the KS2 software suite \citep{Anderson08}.  This photometry
was used to construct color-magnitude diagrams (CMDs) in (\br, \R) and
(\hr, \R).  A search was made of the region around each of the 39
\chandra\ sources within the half-mass radius to locate
potential optical counterparts based on CMD location.  The search radius was
chosen to be the larger of 2.5 times the formal \emph{pwdetect} error
circle radius and 0.3\arcsec.  This choice is motivated by the
observation of \citet{Hong05} that wavelet detection algorithms
systematically underestimate positional uncertainty.  Their
prescription for determining positional uncertainty produces an
asymptotic lower-limiting value of about 0.3\arcsec\ for an on-axis
source.

\begin{figure}
\includegraphics[width=\columnwidth]{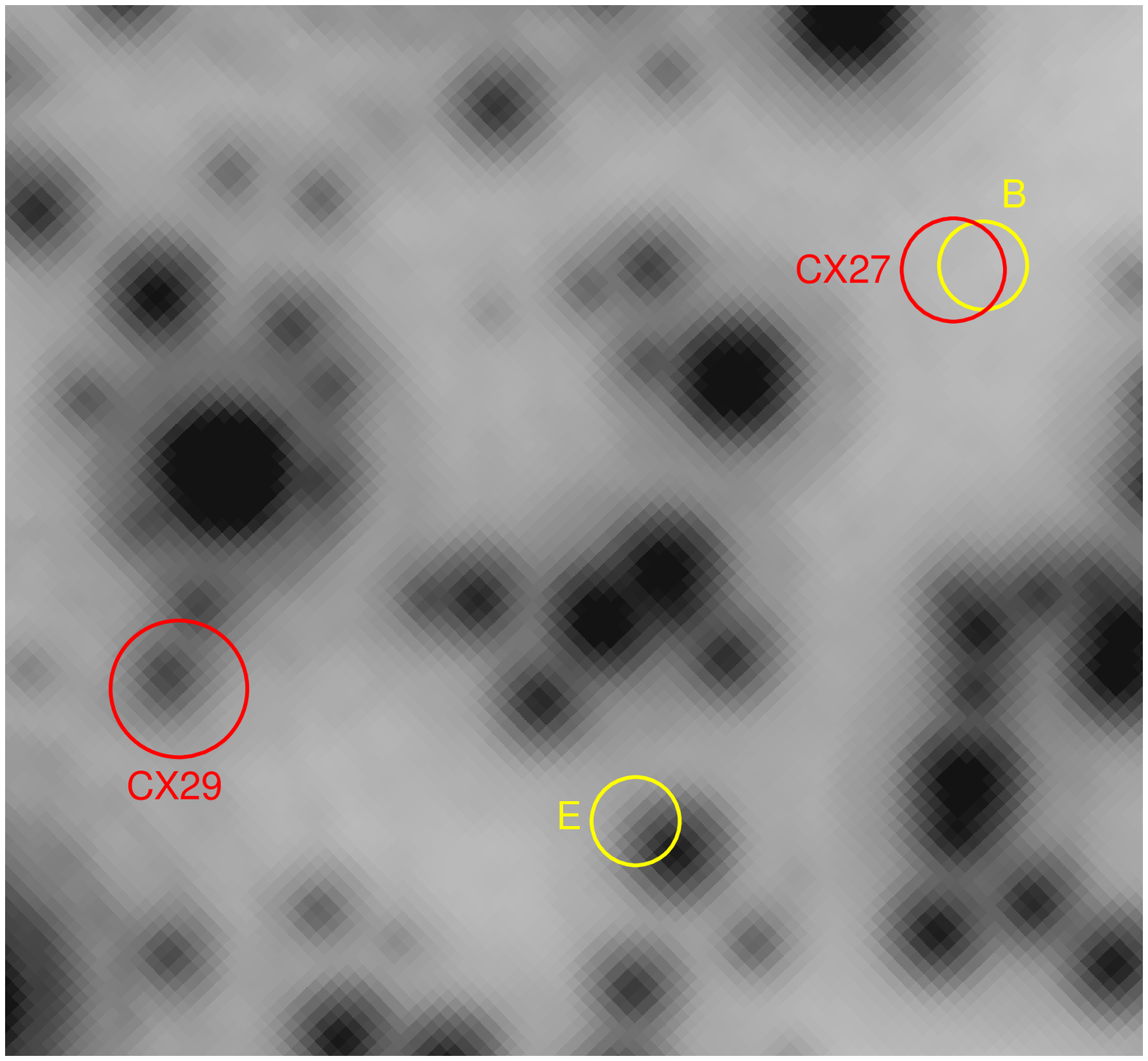}
\includegraphics[width=\columnwidth]{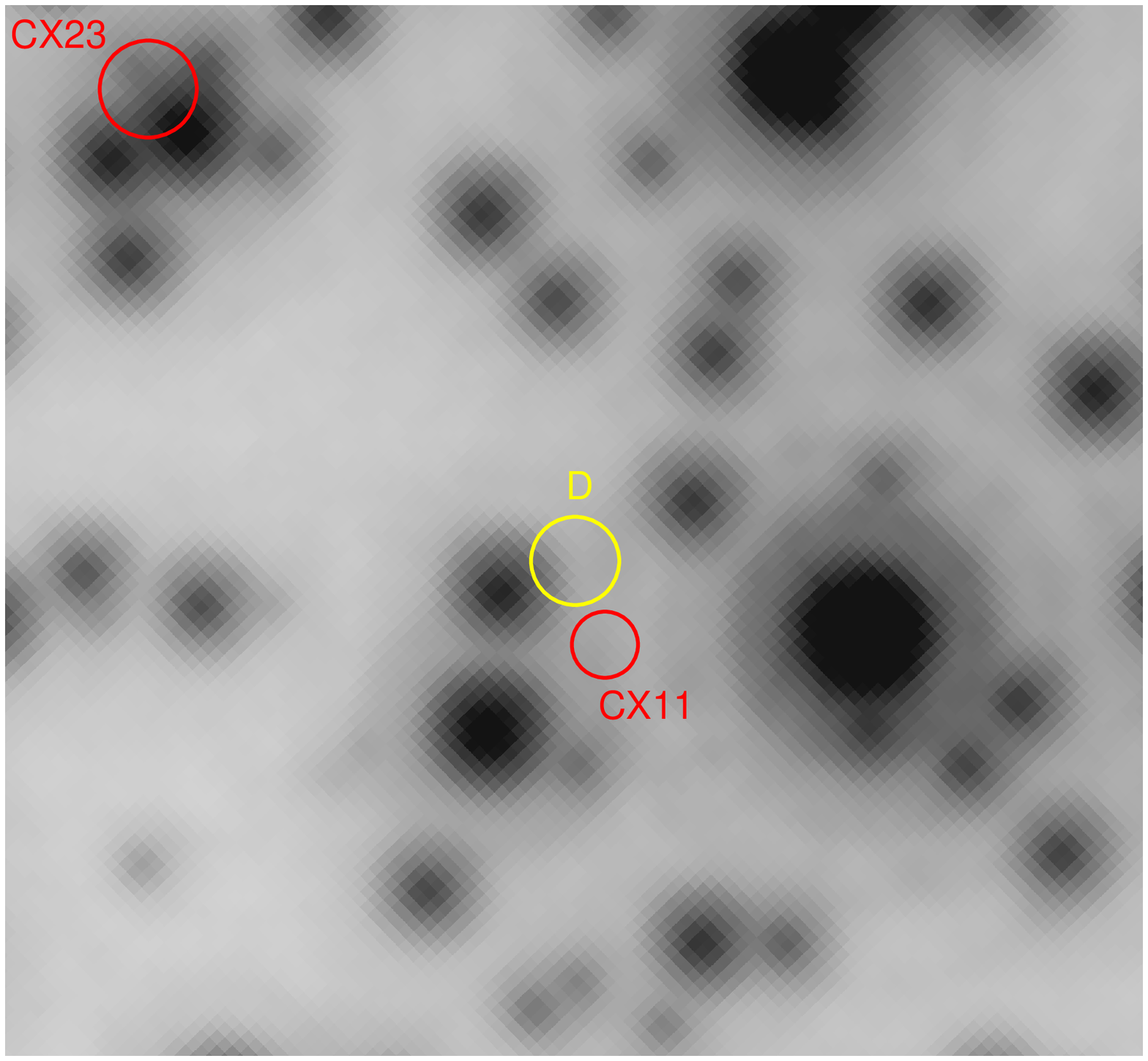}
\caption{Finding charts for MSPs B, D, and E from the drizzled
  \ha\ mosaic.  The displayed regions are 2.4\arcsec\ on a side with a
  $2\times$ oversampled pixel size of 0.025\arcsec.  N is up and E is
  on the left.  The lettered (yellow) positional uncertainty circles
  correspond to the MSPs, while the numbered (red) positional
  uncertainty circles correspond to nearby \chandra\ sources.  Note
  the close positional coincidences between MSP B and \chandra\ source
  {\it CX27} and between MSP D and \chandra\ source {\it CX11}.}
\label{optical_HST}
\end{figure}

A search was also conducted of the regions around the three MSPs from
\citet{D'Amico02} that fall within the ACS/WFC mosaics,  MSPs B,
D, and E\@.  A search radius of 0.1\arcsec was selected, which is the
quadratic sum of the RMS residual for the ACS/WFC mosaic plate
solution (0.09\arcsec) and the uncertainty of the MSP positions
(0.03\arcsec).  The regions around the MSP locations are shown in
Figure~7.  Figure~7a shows the locations of MSPs B and E, while Figure~7b
shows the location of MSP D\@.  
As seen in Figure~7, there are no candidate
counterparts within either the positional uncertainty circle of MSP B
or within the search area around source CX27, which has a radius of 2.5
times the indicated \emph{pwdetect} positional uncertainty radius.

Similarly, Fig.~7b shows that there is an object near the edge of the positional uncertainty circle (0.1'') of MSP D\@.  Examination of the CMDs indicates that this object is also a
star at the MSTO, in this case without even a hint of an \ha\ excess.
We likewise judge this object to be a chance superposition. 
We do not see the optical counterpart suggested by \citet{Pooley02a} in the CX11 error circle (nor do \citealt{Thomson12}, who conducted their own optical counterpart search).  This does not prove that this object is spurious, since our new data is not substantially deeper (though it is of higher resolution), and the object could be of variable brightness.  If Pooley et al's suggested optical counterpart does contribute X-ray emission to CX11, the X-ray luminosity that we infer here for MSP D should be considered an upper limit.

The X-ray emission at MSP E's position, while it appears consistent with an X-ray source, is not identified as a source by our detection algorithms, due to crowding.  
However, the morphology of the six X-ray photons around its position seems consistent 
with an X-ray source at the position of MSP E, if another source (CX29) lies 1.1'' to its east, as suggested by our detection algorithms.
We expect no optical counterpart for MSP E (which has no binary companion), but we do expect an optical counterpart at the position of CX29 if we have correctly assigned the X-ray flux. 
 Indeed, we do find that CX29 has a likely counterpart showing blue colors and a marginal H$\alpha$ excess, indicative of a candidate CV, within the \chandra\ error circle (Lugger et al. 2013, in prep.).
We only identify one object, at 0.1'' separation, within a 0.25'' circle around the position of MSP E that we search for candidates to produce X-ray flux at that position.  Examination of the (\br, \R) CMD indicates that this star 
lies at the main-sequence turnoff (MSTO), while examination of the
(\hr, \R) CMD indicates that it has at best a hint of an \ha\ excess.
The density of stars in this region (shown in Figure 7) predicts 1.7 stars (on average) within a 0.25'' radius circle, or a 1/4 chance of a star within 0.1''.
We therefore judge this object to be a chance superposition of a normal MSTO
star with the MSP E position.  

Thus, we
find no likely optical counterparts for any of MSPs B, D, and E, in agreement with their lack of binary companions (vs. faint WD companions detected in clusters, e.g. MSP A in NGC 6752, \citealt{Bassa03}, \citealt{Ferraro03}, and 47 Tuc U, \citealt{Edmonds01}).  We conclude that \chandra\ sources CX27 and CX11 are indeed produced by MSPs B and D, and that the X-ray emission at the location of MSP E is probably produced by that MSP. Our non-identification of plausible optical counterparts to X-ray sources near the MSP positions is consistent with the results of a similar search in WFC3 data by \citet{Thomson12}.

\section{Discussion}
The NGC 6752 MSPs appear to have unusually low X-ray luminosities, but high temperatures, when compared to the populations of MSPs observed in the other nearby globular clusters 47 Tuc, NGC 6397, M28, M4, and M71 (see \citealt{Bogdanov06,Bogdanov10,Bogdanov11,Bassa04,Elsner08}).  This cannot be attributed to differences in sensitivity, since our observations do not reach to as low X-ray luminosities as those of 47 Tuc, NGC 6397, or M4.
Below we consider whether we can identify clear variations in either luminosity or temperature, and whether there may be an obvious explanation if so.

\begin{figure}
\includegraphics[width=\columnwidth]{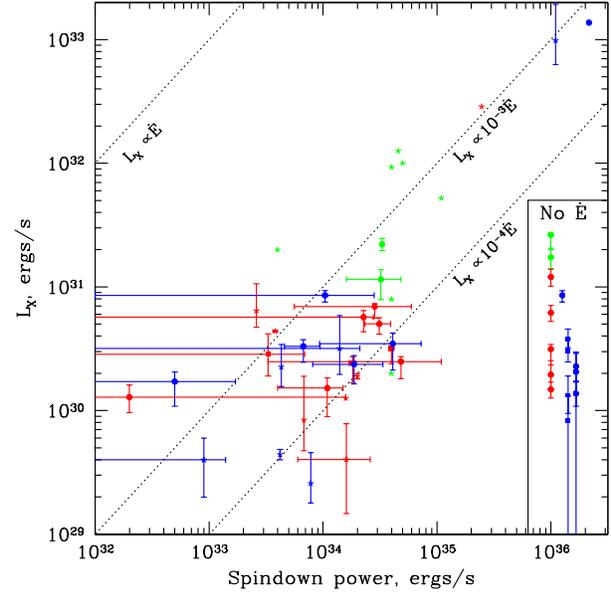}
\caption{Reported values of $L_X$ vs. spindown power for radio MSPs, from this work and the literature (Table 4). Red symbols indicate binary MSPs, blue symbols indicate solitary MSPs, and green symbols indicate binary MSPs with evidence for X-ray emission from an intrabinary wind shock.  Filled circles indicate MSPs in globular clusters, while asterisks indicate MSPs outside clusters.  MSPs without reliable spindown power measurements have their $L_X$ plotted in the small box on the right.}
\label{lxvEdot}
\end{figure}

Kolmogorov-Smirnov tests comparing the luminosity distributions of the NGC 6752 MSPs with either the 47 Tuc MSPs, or to the MSPs in all clusters listed above, indicate a probability $>$10\% of obtaining this result by chance.  Thus we quickly conclude that there is no evidence that the $L_X$ values of the NGC 6752 MSPs are unusual.  However, this draws our attention to another possibility. 
Several very X-ray faint ($L_X\simle 10^{30}$ ergs/s) MSPs, in both the field and globular clusters, are isolated; PSR B1257+12 \citep[which has planets, but no companion, so is considered isolated][]{Pavlov07}, PSR J1024-0719, \citep{Becker99}; PSR J1744-1134, \citep{Kargaltsev12} ; and now NGC 6752 E.  This is of particular interest given recent evidence that the radio luminosities of binary and isolated recycled pulsars differ \citep{Burgay13}.  

We have compiled estimates of the X-ray luminosity (in the 0.3-8 keV band, as this corresponds reasonably to what can actually be measured) for MSPs (pulsars with P$<$20 ms) both in clusters and the field, in Table 4.  We include errors on the fluxes and distances (in many cases from parallax measurements); the distance errors typically dominate $L_X$ uncertainties for field MSPs, while the flux measurements dominate uncertainties in $L_X$ for globular cluster MSPs.  We include spindown luminosities where possible, and plot $L_X$ vs. spindown power (with errors where calculated) in Figure 8.  It is clear that, although there are more X-ray faint isolated MSPs than X-ray faint MSPs in binaries, there is not a significant statistical difference between the thermal $L_X$ of the two populations.  Ignoring the three MSPs with high spindown energy, and those binary MSPs showing evidence (typically from orbital variability) for a shocked intrabinary wind producing the majority of X-rays \citep[e.g.,][]{Bogdanov05}, the binary and isolated MSPs have consistent distributions.  A Kolmogorov-Smirnov test gives a probability $>$10\% of measuring such a difference even if the two groups have the same parent distribution.  There is also no evidence for a difference in the spindown power distributions of binary vs. isolated MSPs, or for a difference in the relation of $L_X$ and spindown power for the two groups.

The best fit spectral models of the NGC 6752 MSPs predict generally higher temperatures than seen in the other clusters (Figure~\ref{Lx_T}).  (Note that the NSATMOS hydrogen atmosphere model gives lower estimates of the temperatures (Table \ref{BB_NSATMOS}), while the NSATMOS unabsorbed luminosity estimates agree with those from the BB model. )
Unlike for the X-ray luminosities, here we identify a statistically significant difference. 
A Kolmogorov-Smirnov test, between the inferred blackbody temperatures of the NGC 6752 MSPs and those of the 47 Tuc MSPs, gives a $<$1\% probability of obtaining such dramatically different samples if the parent temperature distributions were identical. This temperature difference, combined with the similar or smaller luminosities, suggests that the emitting regions of the MSPs in NGC 6752 are smaller.  Modelling the effective radius and temperature simultaneously in XSPEC (Figure~\ref{T_R}), we confirm that smaller effective emitting radii are required to model the NGC 6752 MSPs.

\begin{figure}
\includegraphics[width=\columnwidth]{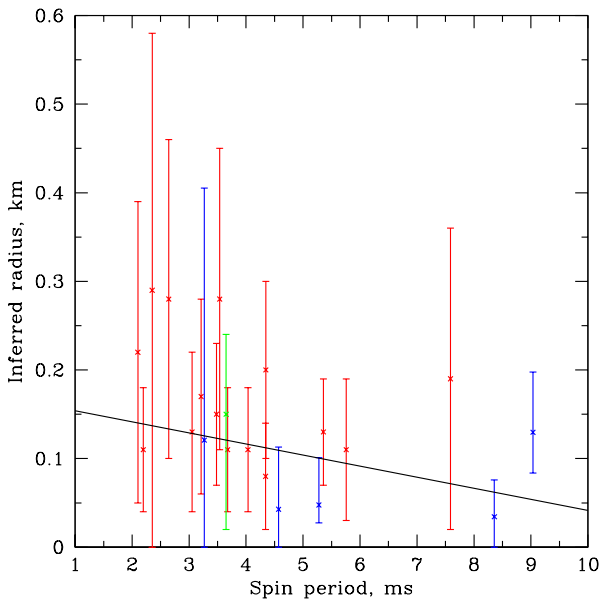}
\caption{Fitted (blackbody) polar cap radius vs. spin period for radio MSPs in NGC 6752 (blue), NGC 6397 (green), and 47 Tuc (red).  The best-fitting power-law is indicated, with a best-fit slope of -0.65$\pm0.40$, consistent with the theoretically predicted -0.5.}
\label{spin_cap}
\end{figure}

The high modeled temperatures of the NGC 6752 MSPs could be due to X-ray source confusion in the region (photons from higher-temperature sources nearby biasing the temperature estimates), or to magnetospheric emission from these MSPs (a high-energy power-law component, which cannot be identified with these low-statistic spectra).  More interestingly, the predicted size of the polar cap region $R_{\rm pc}=(2\pi R_{\rm NS}/(cP))^{1/2} R_{\rm NS}$ (e.g. \citealt{Lyne06}) depends inversely on the spin period.  Since the NGC 6752 MSPs have longer periods on average than the 47 Tuc MSPs, there is thus a clear rationale for them to have smaller polar caps and (given similar luminosities) relatively higher polar cap temperatures.  To test this idea, we plot inferred MSP effective radii (from single-temperature blackbody fits) vs. spin periods for the MSPs in 47 Tuc, NGC 6397, and NGC 6752 (Figure~\ref{spin_cap}), which suggests a correlation.  Fitting the effective radii measurements with a power-law in spin period, we find a best-fit index of -0.65$\pm0.40$ (1$\sigma$ errorbars), which is indeed consistent with the predicted index of -0.5 (though it has rather large errorbars).  This correlation could easily be weakened by the (unknown) differences in geometries of the pulsars, and by variations in the strength of unmodeled nonthermal radiation.  Nevertheless, following this suggested correlation up with detailed analyses of high-quality archival X-ray spectra of nearby MSPs, and deeper observations of globular cluster MSP populations (including NGC 6752 and 47 Tuc), might verify this long-predicted relation.

\section*{Acknowledgements}

COH and GRS are supported by NSERC Discovery Grants. COH is also supported by an Ingenuity New Faculty Award, and LMF is supported by an NSERC USRA award. 
This work includes observations made with the NASA/ESA Hubble Space Telescope. AC, HC, and PL acknowledge support from NASA grant HST-GO-12254 from the Space Telescope Science Institute, which is operated by the Association of Universities for Research in Astronomy, Incorporated, under NASA contract NAS5-26555. 
This research has made use of the NASA Astrophysics Data System (ADS)
and software provided by the Chandra X-ray Center (CXC) in the
application package CIAO. 
Pwdetect has been developed by scientists at Osservatorio Astronomico di Palermo G. S. Vaiana thanks to Italian CNAA and MURST (COFIN) grants. 


\bibliographystyle{mn2e}
\bibliography{odd_src_ref_list}

\clearpage

{\small
\renewcommand{\arraystretch}{1.5}
\onecolumn
\begin{longtable}{c c c  |  c c c | c c}
\caption{Basic X-Ray Properties of Catalog Sources in NGC 6752}\\
\hline
\hline
 &  \multicolumn{2}{c}{Position (J2000)}  & \multicolumn{3}{c}{Counts} & $L_X$ & \\
NAME & \bfalpha & \bfdelta &  0.5-1.5 & 1.5-6 & 0.5-8 & 0.5-8 keV & TYPE \\ 
 &  \phantom{X} (h m s, \arcsec err) \phantom{X} & \phantom{X}  (\deg \arcmin \arcsec, \arcsec err) \phantom{X}  & \phantom{X}  keV \phantom{X} & \phantom{X}  keV \phantom{X}  & \phantom{X}  keV \phantom{X}  & (ergs $s^{-1}$) &  \\
\hline
\hline
\endhead
\hline
\endfoot
\multicolumn{8}{p{\textwidth}}{Note: Catalog source parameters derived using XSPEC. The source positions are adjusted to place them onto the radio frame, using the X-ray detections of three radio MSPs.  Positional errors are quoted in arcseconds for both RA and Dec, and include only statistical errors.  The nine brightest sources were fitted individually with BREMSS models, while the combined faint source spectrum was fitted using a MEKAL model. The catalog sources corresponding to MSPs were fitted with the XSPEC BB model.}\\
\endlastfoot
CX1  &  19 10 51.135 0.009  &  -59 59 11.745 0.010  & $578.8^{+25.1}_{-24.1}$ & $494.9^{+23.3}_{-22.2}$ & $1092.6^{+34.0}_{-33.0}$ & $273.5^{+18.4}_{-18.4}$   & CV\\
CX3  &  19 10 40.375 0.011  &  -59 58 41.345 0.015  & $433.8^{+21.9}_{-20.8}$ & $268.8^{+17.4}_{-16.4}$ & $707.5^{+27.6}_{-26.6}$  & $145.2^{+13.3}_{-13.1}$   & CV\\
CX2  &  19 10 56.005 0.014  &  -59 59 37.245 0.013  & $344.9^{+19.6}_{-18.6}$ & $245.8^{+16.7}_{-15.7}$ & $597.6^{+25.4}_{-24.4}$  & $133.1^{+12.7}_{-12.6}$   & CV\\
CX5  &  19 10 51.415 0.015  &  -59 59 05.045 0.015  & $184.7^{+14.6}_{-13.6}$ & $190.7^{+14.8}_{-13.8}$ & $376.3^{+20.4}_{-19.4}$  & $107.0^{+13.1}_{-12.7}$   & CV/BY?\\
CX4  &  19 10 51.585 0.015  &  -59 59 01.645 0.016  & $181.6^{+14.5}_{-13.5}$ & $157.6^{+13.6}_{-12.6}$ & $342.1^{+19.5}_{-18.5}$  & $93.3^{+12.1}_{-11.8}$    & CV \\
CX7  &  19 10 51.505 0.016  &  -59 58 56.745 0.016  & $174.8^{+14.3}_{-13.2}$ & $133.8^{+12.6}_{-11.6}$ & $314.5^{+18.7}_{-17.7}$  & $72.3^{+9.9}_{-9.7}$      & CV \\
CX8  &  19 11 02.965 0.040  &  -59 59 41.745 0.029  & $115.9^{+11.8}_{-10.8}$ & $7.8^{+4.0}_{-2.8}$     & $124.6^{+12.2}_{-11.2}$  & $49.9^{+173.5}_{-29.6}$   & -- \\
CX6  &  19 10 51.505 0.020  &  -59 59 26.945 0.021  & $156.8^{+13.6}_{-12.5}$ & $72.8^{+9.6}_{-8.5}$    & $232.5^{+16.3}_{-15.3}$  & $44.6^{+6.9}_{-6.6}$      & CV \\
CX9  &  19 10 51.765 0.025  &  -59 58 59.145 0.025  & $91.1^{+10.6}_{-9.6}$   & $34.3^{+7.0}_{-5.9}$    & $126.4^{+12.3}_{-11.3}$  & $22.6^{+6.6}_{-5.3}$      & -- \\
CX10 &  19 10 54.755 0.044  &  -59 59 13.745 0.043  & $24.8^{+6.1}_{-5.0}$    & $21.8^{+5.8}_{-4.7}$    & $46.6^{+7.9}_{-6.8}$     & $9.2^{+1.6}_{-1.3}$       & CV \\
CX13 &  19 10 40.605 0.083  &  -60 00 05.745 0.110  & $11.8^{+4.6}_{-3.4}$    & $16.8^{+5.2}_{-4.1}$    & $28.4^{+6.5}_{-5.4}$     & $5.6^{+1.3}_{-1.1}$       & CV \\
CX14 &  19 10 52.065 0.055  &  -59 59 08.945 0.057  & $24.8^{+6.1}_{-5.0}$    & $2.8^{+2.9}_{-1.6}$     & $27.5^{+6.4}_{-5.3}$     & $5.4^{+1.3}_{-1.0}$       & - \\
CX15 &  19 10 55.845 0.058  &  -59 57 45.645 0.061  & $15.8^{+5.1}_{-4.0}$    & $6.8^{+3.8}_{-2.6}$     & $22.6^{+5.9}_{-4.8}$     & $4.4^{+1.2}_{-0.9}$       & CV? \\
CX11 &  19 10 52.405 0.052  &  -59 59 05.545 0.054  & $29.6^{+6.5}_{-5.4}$    & $1.7^{+2.7}_{-1.3}$     & $32.3^{+6.8}_{-5.7}$     & $3.6^{+0.7}_{-0.6}$       & MSP D\\
CX20 &  19 10 52.845 0.069  &  -59 59 02.445 0.070  & $12.7^{+4.7}_{-3.6}$    & $3.8^{+3.2}_{-1.9}$     & $16.4^{+5.2}_{-4.1}$     & $3.2^{+1.0}_{-0.8}$       & -- \\
CX21 &  19 10 49.515 0.082  &  -59 59 43.045 0.069  & $9.8^{+4.3}_{-3.1}$     & $5.8^{+3.6}_{-2.4}$     & $15.5^{+5.0}_{-3.8}$     & $3.1^{+1.0}_{-0.8}$       & -- \\
CX22 &  19 11 2.945 0.095  &  -59 57 58.745 0.092  & $3.8^{+3.2}_{-1.9}$     & $9.8^{+4.3}_{-3.1}$     & $13.5^{+4.8}_{-3.7}$     & $2.7^{+1.0}_{-0.7}$       & -- \\
CX16 &  19 10 42.535 0.078  &  -59 58 42.945 0.107  & $12.9^{+4.7}_{-3.6}$    & $0.8^{+2.3}_{-0.8}$     & $13.7^{+4.8}_{-3.7}$     & $2.7^{+1.0}_{-0.7}$       & -- \\
CX23 &  19 10 52.545 0.078  &  -59 59 04.245 0.078  & $11.3^{+4.6}_{-3.4}$    & $0.7^{+2.3}_{-0.8}$     & $11.9^{+4.7}_{-3.6}$     & $2.3^{+0.9}_{-0.7}$       & -- \\
CX24 &  19 10 52.665 0.084  &  -59 59 03.045 0.088  & $8.4^{+4.1}_{-2.9}$     & $2.7^{+2.9}_{-1.6}$     & $11.1^{+4.6}_{-3.4}$     & $2.2^{+0.9}_{-0.7}$       & -- \\
CX17 &  19 11 05.255 0.128  &  -59 59 04.245 0.096  & $3.9^{+3.2}_{-1.9}$     & $6.8^{+3.8}_{-2.6}$     & $10.6^{+4.4}_{-3.3}$     & $2.1^{+0.9}_{-0.6}$       & -- \\
CX25 &  19 10 51.955 0.092  &  -59 58 40.445 0.093  & $3.9^{+3.2}_{-1.9}$     & $4.8^{+3.4}_{-2.2}$     & $8.7^{+4.1}_{-2.9}$      & $1.7^{+0.8}_{-0.6}$       & -- \\
CX18 &  19 10 52.055 0.099  &  -59 59 03.545 0.105  & $7.5^{+4.0}_{-2.8}$     & $0.7^{+2.3}_{-0.8}$     & $8.1^{+4.1}_{-2.9}$      & $1.6^{+0.8}_{-0.6}$       & -- \\
CX19 &  19 10 55.595 0.114  &  -59 59 17.245 0.109  & $5.8^{+3.6}_{-2.4}$     & $1.8^{+2.7}_{-1.3}$     & $7.6^{+4.0}_{-2.8}$      & $1.5^{+0.8}_{-0.5}$       & -- \\
CX26 &  19 10 39.165 0.144  &  -59 59 45.045 0.178  & $2.8^{+2.9}_{-1.6}$     & $4.8^{+3.4}_{-2.2}$     & $7.5^{+4.0}_{-2.8}$      & $1.5^{+0.8}_{-0.5}$       & -- \\
CX27 &  19 10 52.055 0.082  &  -59 59 00.745 0.083  & $8.4^{+4.1}_{-2.9}$     & $1.6^{+2.7}_{-1.3}$     & $9.9^{+4.4}_{-3.3}$      & $1.3^{+0.5}_{-0.4}$       & MSP B\\
CX28 &  19 10 42.505 0.141  &  -59 59 44.345 0.171  & $6.9^{+3.8}_{-2.6}$     & $-0.2^{+1.9}_{-0}$      & $6.5^{+3.8}_{-2.6}$      & $1.3^{+0.7}_{-0.5}$       & -- \\
CX29 &  19 10 52.295 0.109  &  -59 59 01.645 0.110  & $4.6^{+3.4}_{-2.2}$     & $1.8^{+2.7}_{-1.3}$     & $6.4^{+3.8}_{-2.6}$      & $1.3^{+0.7}_{-0.5}$       & -- \\
CX30 &  19 10 40.675 0.113  &  -59 58 39.445 0.152  & $3.3^{+3.2}_{-1.9}$     & $2.4^{+2.9}_{-1.6}$     & $5.6^{+3.8}_{-2.6}$      & $1.1^{+0.7}_{-0.5}$       & -- \\
CX31 &  19 10 50.515 0.112  &  -59 57 36.945 0.124  & $4.9^{+3.4}_{-2.2}$     & $0.9^{+2.3}_{-0.8}$     & $5.6^{+3.6}_{-2.4}$      & $1.1^{+0.7}_{-0.5}$       & -- \\
CX32 &  19 10 54.135 0.123  &  -59 59 10.945 0.115  & $1.9^{+2.7}_{-1.3}$     & $3.8^{+3.2}_{-1.9}$     & $5.6^{+3.6}_{-2.4}$      & $1.1^{+0.7}_{-0.5}$       & -- \\
CX33 &  19 11 03.285 0.141  &  -59 58 01.145 0.133  & $3.8^{+3.2}_{-1.9}$     & $0.8^{+2.3}_{-0.8}$     & $5.5^{+3.6}_{-2.4}$      & $1.1^{+0.7}_{-0.5}$       & -- \\
CX34 &  19 10 45.695 0.125  &  -59 58 19.945 0.156  & $3.9^{+3.2}_{-1.9}$     & $0.9^{+2.3}_{-0.8}$     & $4.7^{+3.4}_{-2.2}$      & $0.9^{+0.7}_{-0.4}$       & -- \\
CX35 &  19 10 52.165 0.136  &  -59 59 16.645 0.132  & $0.8^{+2.3}_{-0.8}$     & $3.9^{+3.2}_{-1.9}$     & $4.6^{+3.2}_{-1.9}$      & $0.9^{+0.6}_{-0.4}$       & -- \\
CX36 &  19 10 49.585 0.139  &  -59 58 26.245 0.148  & $2.9^{+2.9}_{-1.6}$     & $0.9^{+2.3}_{-0.8}$     & $3.7^{+3.2}_{-1.9}$      & $0.7^{+0.6}_{-0.4}$       & -- \\
CX37 &  19 10 50.505 0.148  &  -59 59 08.645 0.151  & $2.9^{+2.9}_{-1.6}$     & $0.9^{+2.3}_{-0.8}$     & $3.7^{+2.9}_{-1.6}$      & $0.7^{+0.6}_{-0.3}$       & -- \\
CX38 &  19 11 02.155 0.173  &  -59 58 11.645 0.162  & $2.9^{+2.9}_{-1.6}$     & $0.9^{+2.3}_{-0.8}$     & $3.7^{+3.2}_{-1.9}$      & $0.7^{+0.6}_{-0.4}$       & -- \\
CX39 &  19 10 46.355 0.139  &  -59 57 49.745 0.168  & $3.9^{+3.2}_{-1.9}$     & $-0.2^{+1.9}_{-0}$      & $3.6^{+3.2}_{-1.9}$      & $0.7^{+0.6}_{-0.4}$       & -- \\
CX40 &  19 10 50.355 0.183  &  -59 59 13.745 0.203  & $2.9^{+2.9}_{-1.6}$     & $-0.1^{+1.9}_{-0}$      & $2.6^{+2.9}_{-1.6}$      & $0.5^{+0.6}_{-0.3}$       & -- \\
\label{Catalog_Parameters}
\end{longtable}}

{\small
\renewcommand{\arraystretch}{1.5}
\begin{longtable}{p{0.05\textwidth}  p{0.08\textwidth} p{0.08\textwidth} | p{0.08\textwidth} p{0.08\textwidth} p{0.08\textwidth} }
\caption{Radio Positions and X-ray Countrates of the MSPs in NGC 6752}\\
\hline\hline
    & \multicolumn{2}{c}{Position (J2000)}             &  &Counts &       \\
MSP & \bfalpha & \bfdelta                                & 0.5-1.5 & 1.5-6 & 0.5-8 \\ 
    & (h m s)  & (\deg \arcmin \arcsec)                  &  keV    &  keV  &  keV \\
\hline

A$^*$ & 19 11 42.76 & -59 58 26.9 &  \phantom{-}$9.9^{+4.6}_{-3.4}$ & $-0.7^{+3.0}_{-1.7}$ & $13.9^{+5.5}_{-4.4}$ \\
B     & 19 10 52.05 & -59 59 00.8 &  \phantom{-}$8.4^{+4.1}_{-2.9}$ & \phantom{-}$1.6^{+2.7}_{-1.3} $ & \phantom{-}$9.9^{+4.4}_{-3.3}$  \\	
C     & 19 11 05.56 & -60 00 59.7 &  $18.7^{+5.4}_{-4.3}$& \phantom{-}$4.6^{+3.4}_{-2.2}$  & $23.0^{+6.0}_{-4.9}$ \\
D     & 19 10 52.42 & -59 59 05.5 &  $29.6^{+6.5}_{-5.4}$& \phantom{-}$1.7^{+2.7}_{-1.3}$  & $32.3^{+6.8}_{-5.7}$ \\
E     & 19 10 52.16 & -59 59 02.1 &  \phantom{-}$2.9^{+2.9}_{-1.6}$ & \phantom{-}$0.9^{+2.3}_{-0.8}$  & \phantom{-}$3.7^{+3.2}_{-1.9}$  \\
\hline
\multicolumn{6}{p{0.55\textwidth}}{Note:: $^*$ MSP A was only observed in the 6612 observation.  Radio positions from \citet{D'Amico02}.}\\
\label{MSP_Parameters}
\end{longtable}
}


{\small
\renewcommand{\arraystretch}{1.5}
\begin{longtable}{p{0.12\textwidth} p{0.12\textwidth} p{0.12\textwidth} p{0.12\textwidth} p{0.12\textwidth} p{0.12\textwidth}}
\caption{X-Ray Spectral Properties of the MSPs in NGC 6752}\\
\hline\hline
MSP & Model & kT    & $R_{\rm eff}$   & $L_X$                   & goodness \\
    &       & (keV) & (km)       & ($10^{30}$ergs $s^{-1}$) & (\%)     \\
\hline
A$^*$  & BB       &  $0.21^{+0.10}_{-0.06}$ & $0.12^{+0.28}_{-0.12}$ & $2.5^{+1.1}_{-0.9}$  & 75.47 \\
       & NSATMOS  &  $0.11^{+0.19}_{-0.06}$ & $0.58^{+4.91}_{-0.58}$  & $2.6^{+188}_{-2.6}$  & 99.57 \\
\hline
B      & BB       &  $0.31^{+0.10}_{-0.06}$ & $0.03^{+0.04}_{-0.03}$ & $1.3^{+0.5}_{-0.4}$  & 50.31 \\
       & NSATMOS  &  $0.24^{+0.21}_{-0.11}$ & $0.08^{+0.18}_{-0.08}$ & $1.3^{+17}_{-1.2}$  & 64.85 \\
\hline
C      & BB       &  $0.31^{+0.06}_{-0.06}$ & $0.05^{+0.05}_{-0.02}$ & $2.7^{+0.7}_{-0.6}$  & 58.56 \\
       & NSATMOS  &  $0.20^{+0.10}_{-0.07}$ & $0.16^{+0.25}_{-0.16}$ & $2.8^{+12}_{-2.4}$  & 44.63 \\
\hline
D      & BB       &  $0.21^{+0.03}_{-0.02}$ & $0.13^{+0.07}_{-0.05}$ & $3.4^{+0.6}_{-0.6}$  & 30.55 \\
       & NSATMOS  &  $0.13^{+0.04}_{-0.04}$ & $0.52^{+0.55}_{-0.52}$ & $3.4^{+12}_{-2.7}$  & 42.56 \\
\hline   
E      & BB       &  $0.27^{+0.19}_{-0.09}$ & $0.04^{+0.07}_{-0.04}$ & $1.0^{+0.9}_{-0.5}$  & 54.04 \\
       & NSATMOS  &  $0.18^{+0.21}_{-0.09}$ & $0.13^{+0.57}_{-0.13}$ & $1.0^{+27}_{-1.0}$  & 52.30 \\
\hline
\multicolumn{6}{p{0.80\textwidth}}{Note:: name$^*$ indicates the source was only observed in the 6612 observation. Both models were fitted using XSPEC, modified by TBABS. The BB model used CFLUX to determine the unabsorbed luminosity, while the NSATMOS model used the XSPEC FLUX command to estimate the unabsorbed (0.5-8 keV) luminosity. }\\
\label{BB_NSATMOS}
\end{longtable}
}

\clearpage 
{\small
\renewcommand{\arraystretch}{1.5}
\begin{longtable}{p{0.12\textwidth} p{0.12\textwidth} p{0.12\textwidth} p{0.12\textwidth} p{0.12\textwidth} p{0.12\textwidth}}
\caption{Properties of X-ray Detected MSPs}\\
\hline\hline
MSP & \.{E}         & $F_X (0.3-8 keV)$        & Dist    & Nature$^a$    & Refs \\
  & ($10^{34}$ ergs/s) & ($10^{-15}$ ergs s$^{-1}$ cm$^{-2}$) & (kpc)  & &   \\
\hline
\multicolumn{6}{c}{Globular Cluster Sources} \\
\hline
NGC 6752A & 0.33$^{+0.36}_{-0.33}$ & 1.5$^{+0.7}_{-0.5}$ & 4.0 & B & 1,7,8,28 \\ 
NGC 6752B & - & 0.7$^{+0.3}_{-0.2}$ & 4.0 & I & 1,7,28\\ 
NGC 6752C & - & 1.65$^{+0.4}_{-0.3}$ & 4.0 & I & 1,7,8,28\\ 
NGC 6752D & - & 2.0$^{+0.4}_{-0.3}$ & 4.0 & I & 1,7,28\\ 
NGC 6752E & - & 0.5$^{+0.5}_{-0.2}$ & 4.0 & I & 1,7,28\\ 
\hline
NGC 104C & 0.05$^{+0.12}_{-0.05}$ & 0.71$^{+0.14}_{-0.26}$ & 4.5 & I & 1,9\\ 
NGC 104D & 0.67$^{+0.27}_{-0.21}$ & 1.37$^{+0.18}_{-0.35}$ & 4.5 & I & 1,9\\ 
NGC 104E & 3.12$^{+0.79}_{-0.79}$ & 2.08$^{+0.26}_{-0.36}$ & 4.5 & B & 1,9\\ 
NGC 104F & 4.09$^{+3.16}_{-3.16}$ & 1.44$^{+0.31}_{-0.56}$ & 4.5 & I & 1,9\\ 
NGC 104H & - & 1.30$^{+0.12}_{-0.33}$ & 4.57 & B & 1,9\\ 
NGC 104J & 3.22$^{+1.61}_{-1.61}$ & 4.77$^{+0.95}_{-1.51}$ & 4.5 & B,S & 1,9\\ 
NGC 104L & 1.04$^{+1.77}_{-1.04}$ & 3.54$^{+0.32}_{-0.41}$ & 4.5 & I & 1,9\\ 
NGC 104M & - & 1.01$^{+0.14}_{-0.30}$ & 4.57 & I & 1,9\\ 
NGC 104N & 1.87$^{+1.49}_{-1.06}$ & 0.98$^{+0.17}_{-0.30}$ & 4.5 & I & 1,9\\ 
NGC 104Q & 1.82$^{+0.12}_{-0.12}$ & 1.00$^{+0.13}_{-0.30}$ & 4.5 & B & 1,9\\ 
NGC 104R & 2.84$^{+3.10}_{-2.28}$ & 2.87$^{+0.17}_{-0.56}$ & 4.5 & B & 1,9\\ 
NGC 104S & 2.27$^{+0.49}_{-2.27}$ & 2.36$^{+0.31}_{-0.56}$ & 4.5 & B & 1,9\\ 
NGC 104T & 1.09$^{+0.41}_{-0.69}$ & 0.63$^{+0.13}_{-0.26}$ & 4.5 & B & 1,9\\ 
NGC 104U & 3.98$^{+0.21}_{-0.21}$ & 1.32$^{+0.12}_{-0.33}$ & 4.5 & B & 1,9\\ 
NGC 104W & - & 10.9$^{+0.40}_{-2.61}$ & 4.5 & B,S & 1,9\\ 
NGC 104Y & 4.82$^{+6.12}_{-4.49}$ & 1.03$^{+0.10}_{-0.28}$ & 4.5 & B & 1,9\\ 
\hline
NGC 6397A & 3.3         & 1.03$^{+0.10}_{-0.28}$ & 2.3 & B & 1,10, 30\\ 
\hline
M4A  & 0.02$^{+1.56}_{-0.02}$ & 3.6$^{+0.9}_{-0.9}$ & 2.2 & B & 1,11,31\\ 
\hline
M28A & 216 & 380$^{+13}_{-9}$ & 5.5 & I & 1,12,32\\ 
M28B & - & 0.57$^{+0.23}_{-0.56}$ & 5.5 & I & 1,13,32\\ 
M28C & - & 0.54$^{+0.16}_{-0.13}$ & 5.5 & B & 1,13,32\\ 
M28E & - & 0.63$^{+0.19}_{-0.16}$ & 5.5 & I & 1,13,32\\ 
M28F & - & 0.38$^{+0.10}_{-0.08}$ & 5.5 & I & 1,13,32\\ 
M28 H & - & 4.8$^{+0.9}_{-3.6}$ & 5.5 & B,S & 1,13,32\\ 
M28 J & - & 0.41$^{+0.06}_{-0.06}$ & 5.5 & B & 1,13,32\\ 
M28 K & - & 1.71$^{+0.25}_{-0.25}$ & 5.5 & B & 1,13,32\\ 
\hline
M71A & - & 6.3$^{+1.0}_{-1.0}$  & 4.0 & B & 1,14,33\\ 
\hline
\multicolumn{6}{c}{Field Sources}\\
\hline
J0437-4715 & 0.38 & 1500$^{+200}_{-300}$ & 0.156$\pm0.001$ & B & 3,15,34\\ 
J0751+1807 & 0.68 & 44 & $0.4^{+0.2}_{-0.1}$ & B & 4,16,35\\ 
J1012+5307 & 0.26 & 1.2$\times10^2$ & $0.7^{+0.2}_{-0.1}$ & B & 4,17,35\\ 
J1909-3744 & 2   & 10 & $1.26^{+0.03}_{-0.03}$ & B & 5,36\\ 
J0218+4232 & 24.6 & 4.2$\times10^2$ & 2.7 & B & 18,37\\ 
B1957+20 & 10.9 & 90 & 2.5 & B,S & 19,38\\ 
J0034-0534 & 3 &  3.0 & 0.53 & B & 20,39\\ 
J0030+0451 & $1.4^{+0.7}_{-1.4}$ & 3.4$\times10^2$ & $0.28^{+0.10}_{-0.06}$ & I & 4,21,40\\ 
J1024-0719 & $0.09^{+0.05}_{-0.09}$ & 20 & $0.39^{+0.04}_{-0.10}$ & I & 2,22,39\\ 
J1744-1134 & 0.42 & 21 & $0.42^{+0.02}_{-0.02}$ & I & 5,22,36\\ 
J2124-3358 & 0.43 & 2.1$\times10^2$ &  $0.30^{+0.07}_{-0.05}$ & I & 22,39\\ 
B1257+12 & 0.78 & 11 & $0.6^{+0.2}_{-0.1}$ & I & 4,23,31\\ 
B1937+21 & 110 & 3.3$\times10^2$ & 3.6 & I & 4,24,41\\ 
J1023+0038 & 4 & $4.15^{+0.15}_{-0.12}$ & $1.37^{+0.04}_{-0.04}$ & B,S & 6,25,42\\ 
J1124-3653 & 0.4 & 58 & 1.7 & B,S & 26,43\\ 
J1810+1744 & 4 & 19 & 1.9 & B,S & 26,43\\ 
J2215+5135 & 5 & 93 & 3.0 & B,S & 26,43\\ 
J2256-1024 & 4 & 46 & 0.6 & B,S & 27,43\\ 
J0023+0923 & 1.6 & 22 & 0.7 & B & 26,43\\ 
J1723-2837 & 4.6 & $1.6\times10^3$ & 0.75 & B,S & 44,45\\ 
\hline
\multicolumn{6}{p{0.80\textwidth}}{
Notes: Spindown powers, unabsorbed X-ray fluxes (measured at infinity), and distances for radio MSPs (P$<$20 ms).  $^a$ B for binary MSP, I for isolated MSP, S for X-ray emission thought due to interbinary shock \citep[e.g.][]{Bogdanov05}. 
We omit MSPs for which we cannot distinguish their X-ray emission from nearby objects; this includes MSPs NGC 104G, I, L, and O, M28 G and L.  We omit M28 I due to its accretion state during deep observations \citep{Papitto13}.  For the faint M28 pulsars (all but A,H), we calculate the errors on their fluxes from Poisson errors on the net counts listed by \citet{Bogdanov11}.
References: Distances: 1:\citet{Harris96}, 2: \citet{Hotan06}, 3: \citet{Deller08}, 4: \citet{Verbiest12}, 5: \citet{Verbiest09}, 6: \citet{Deller12}.  
Timing properties: 7: \citet{D'Amico02}, 8: \citet{Corongiu06}, 9: \citet{Bogdanov06}, 10: \citet{BassaStappers04}; 11: \citet{Lyne88b}, 12: \citet{Lyne87}, 13: \citet{Begin06}, 14: \citet{Hessels07}, 15: \citet{Hotan06}, 16: \citet{Lundgren95}, 17: \citet{Lange01}, 18: \citet{Navarro95}, 19: \citet{Toscano99a}, 20: \citet{Abdo10}, 21: \citet{Lommen00}, 22: \citet{Bailes97}, 23: \citet{Wolszczan00}, 24: \citet{Backer82}, 25: \citet{Archibald09}, 26: \citet{Bangale13}, 27: \citet{Boyles13}.
X-ray properties: 28: this work, 29: \citet{Bogdanov06}, 30: \citet{Bogdanov10}, 31: \citet{Pavlov07}, 32: \citet{Bogdanov11}, 33: \citet{Elsner08}, 34: \citet{Bogdanov13}, 35:  \citet{Webb04a}, 36: \citet{Kargaltsev12}, 37: \citet{Webb04b}, 38: \citet{Stappers03}, 39: \citet{Zavlin06}, 40: \citet{Becker02}, 41: \citet{Nicastro04}, 42: \citet{Bogdanov11b}, 43: \citet{Gentile13}, 44: \citet{Crawford13}, 45: \citet{Bogdanov14}.
 }\\
\label{XrayMSPs}
\end{longtable}
}
\clearpage

\end{document}